\documentclass[twocolumn,showpacs]{revtex4-1}

\pdfoutput=1

\usepackage{amsmath,amssymb}
\usepackage{bm}
\usepackage{graphicx}

\usepackage{color}

\newcommand{\revision}[1]{\textcolor{black}{#1}}

\begin{document}

\title{
Focusing by blocking: \revision{repeatedly generating central density peaks} in self-propelled particle systems by exploiting diffusive processes}

\author{Andreas M.\ Menzel}
\email{menzel@thphy.uni-duesseldorf.de}
\affiliation{
Institut f\"ur Theoretische Physik II: Weiche Materie, Heinrich-Heine-Universit\"at D\"usseldorf, D-40225 D\"usseldorf, Germany
}


\begin{abstract}
Over the past few years the displacement statistics of self-propelled particles has been intensely studied, revealing their long-time diffusive behavior. Here, we demonstrate that a concerted combination of boundary conditions and switching on and off the self-propelling drive can \revision{generate and afterwards arbitrarily often restore} a non-stationary centered peak in their spatial distribution. This corresponds to a partial reversibility of their statistical behavior, in opposition to the above-mentioned long-time diffusive nature. Interestingly, it is a diffusive process that mediates and makes possible \revision{this procedure}. It should be straightforward to verify our predictions in a real experimental system. 
\end{abstract}

\pacs{82.70.Dd,47.63.Gd,87.16.Uv}


\maketitle


\section{Introduction}\label{intro}

Diffusion \cite{fick1855ueber} is one of the most studied and most familiar phenomena in physics. Regular diffusion is well accessible on the different levels of description in statistical mechanics, making it a prominent textbook example \cite{landau1987fluid,kubo1991statistical,zwanzig2001nonequilibrium,kampen2007stochastic}. On the single-particle level, the situation is typically modeled by corresponding Langevin equations \cite{kubo1991statistical,zwanzig2001nonequilibrium,kampen2007stochastic}. These contain a stochastic force term to reproduce the experimentally observed Brownian motion. 
Likewise, the scenario can be characterized by a dynamic equation for the probability density to find a particle at a certain position at a certain time. This type of dynamic equation is usually termed Smoluchowski equation or, more generally, Fokker-Planck equation \cite{kubo1991statistical,risken1996fokker,zwanzig2001nonequilibrium,kampen2007stochastic}. The connections between the particle picture and these continuum equations for the probability density are well established. Finally, on a more macroscopic level, the phenomenon is addressed by monitoring the time evolution of concentration profiles using diffusion equations \cite{landau1987fluid,kubo1991statistical}. 

One particular feature of diffusive phenomena is their irreversibility. In general, it is not observed that an initial density peak that has flattened over time will suddenly rearise. This is reflected by the nomenclature in the theoretical approaches. One refers to the irreversible part of the Fokker-Planck operator \cite{risken1996fokker,kampen2007stochastic} in the probability density picture and to the irreversible currents \cite{kubo1991statistical} in macroscopic descriptions. A deeper theoretical foundation is provided by the H-Theorem \cite{kubo1991statistical,zwanzig2001nonequilibrium,kampen2007stochastic}, which allows a connection to the increase of entropy or decrease in free energy. 

Here, we report on a globally diffusive system that is demonstrated to at least partially break this paradigm of irreversibility. 
Diffusive systems of initially peaked density profiles generally do not reestablish such a density peak at a later time. Nevertheless, the example system analyzed below shows exactly such behavior, at least to some degree.
\revision{There, a non-stationary centered density peak can be generated independently of the initial distribution. Afterwards it can be restored arbitrarily often.}
It is the non-equilibrium nature of the investigated self-propelled particle system that allows for these effects, together with the possibility to switch on and off the self-propulsion.

\section{Model system}\label{model}

Self-propelled particles feature a mechanism of activated motion \cite{ebbens2010pursuit,romanczuk2012active,cates2012diffusive,menzel2015tuned,elgeti2014physics}. On the one hand, this can for example be the crawling or swimming machinery of motile biological microorganisms such as bacteria and algal cells \cite{berg1972chemotaxis,drescher2011fluid,wada2007model,rappel1999self,szabo2006phase, peruani2012collective,wada2013bidirectional}. On the other hand, self-propelled particles were artificially realized for instance in the form of granular hoppers \cite{narayan2007long,kudrolli2008swarming, goohpattader2009experimental, deseigne2010collective} or self-propelled droplets \cite{nagai2005mode,thutupalli2011swarming, yoshinaga2014spontaneous}. A particularly interesting example are Janus particles 
that allow localized heating by light illumination of only one half of their body \cite{jiang2010active,volpe2011microswimmers,buttinoni2012active, palacci2013living,buttinoni2013dynamical}; or they catalyze chemical reactions on only part of their surface \cite{theurkauff2012dynamic,zheng2013non,howse2007self}. The resulting build-up of temperature or concentration gradients along their surfaces induces stress differences that can effectively drive the particles forward in a type of phoretic motion \cite{golestanian2007designing}. As a central common property, the migration direction of self-propelled particles is not imposed or fixed from outside but results from symmetry breaking \cite{menzel2015tuned}. Light-activated self-propulsion offers the advantage of being easily switchable by turning on and off the external illumination \cite{jiang2010active,volpe2011microswimmers,buttinoni2012active, palacci2013living,buttinoni2013dynamical,babel2014swimming}. 

We consider for our purpose mostly the statistics of the motion of non-interacting self-propelled particles. This implies that either dilute systems are investigated, or the averaged statistical behavior of single objects is analyzed. 
At the end we will briefly address the effect of basic steric interactions. 
As is natural e.g.\ for microswimmers, 
we address their overdamped motion. For simplicity and since many experiments are performed in this way \cite{rappel1999self,nagai2005mode,szabo2006phase,narayan2007long, kudrolli2008swarming,goohpattader2009experimental, jiang2010active,thutupalli2011swarming,deseigne2010collective,volpe2011microswimmers, buttinoni2012active,peruani2012collective,theurkauff2012dynamic, zheng2013non,wada2013bidirectional,palacci2013living,buttinoni2013dynamical}, we confine ourselves to two spatial dimensions. The corresponding Langevin equations for a single particle then read (compare e.g.\ Refs.~\cite{hagen2011brownian,zheng2013non})
\begin{eqnarray}
\label{transdiff}
\frac{d\mathbf{r}}{dt} &=& \beta D_t \left[ F \mathbf{\hat{u}} -\nabla V(\mathbf{r}) \right] + \sqrt{2D_t}\,\bm{\xi}_{\mathbf{r}}, \\
\label{rotdiff_phi}
\frac{d\phi}{dt} &=& \sqrt{2D_r}\, \xi_{\phi}.
\end{eqnarray}
Here, $\mathbf{r}$ denotes the particle position and $\mathbf{\hat{u}}$ the orientation of its self-propulsion direction. $F$ sets the strength of self-propulsion that drives the particle forward along $\mathbf{\hat{u}}$. The potential $V(\mathbf{r})$ leads to a force acting only on the particle position. In our case, it will be nonzero only at the system boundaries to include the steric effect of a confining cavity. $t$ denotes time and 
$\beta=(k_BT)^{-1}$ with $k_B$ the Boltzmann constant and $T$ the temperature. In the two-dimensional plane, the self-propulsion direction can be parameterized by a single angle $\phi$ via $\mathbf{\hat{u}}=(\cos\phi,\sin\phi)$. $D_t$ sets the translational and $D_r$ the rotational diffusion constant. Finally, $\bm{\xi}_{\mathbf{r}}$ and ${\xi}_{\phi}$ are Gaussian white noise terms with averages 
$\langle\bm{\xi}_{\mathbf{r}}(t)\rangle=\mathbf{0}$, $\langle\xi_{\phi}(t)\rangle=0$,
$\langle\xi_{\mathbf{r},i}(t)\xi_{\phi}(t')\rangle={0}$, 
$\langle\xi_{\mathbf{r},i}(t)\xi_{\mathbf{r},j}(t')\rangle=\linebreak\delta_{ij}\delta(t-t')$, 
and $\langle\xi_{\phi}(t)\xi_{\phi}(t')\rangle=\delta(t-t')$, 
where $\delta_{ij}$ is the Kronecker delta and $\delta(t-t')$ the Dirac delta function. 


We now switch to the probability density description by deriving from Eqs.~(\ref{transdiff}) and (\ref{rotdiff_phi}) the corresponding Smoluchowski or Fokker-Planck equation \cite{risken1996fokker,zwanzig2001nonequilibrium,sevilla2014theory}. 
In dimensionless units, this dynamic equation for the probability density $\psi(\mathbf{r},\phi,t)$ to find a particle at time $t$ and position $\mathbf{r}$ with orientation $\phi$ of the self-propulsion direction becomes
\begin{eqnarray}
\partial_t\psi & = & 
{}-v_0\left(\cos\phi\:\partial_x\psi+\sin\phi\:\partial_y\psi\right) + \nabla^2\psi + \partial_{\phi}^2\psi \nonumber\\
&&{}+ \nabla\cdot\left[(\nabla V)\psi\right]. 
\label{fp}
\end{eqnarray}
Here, we rescaled time, space, and the potential by\linebreak $t'=tD_r$, $\mathbf{r}'=\mathbf{r}\sqrt{D_r/D_t}$, and $V'=\beta V$, respectively, with the primes omitted in the equation, and we confine ourselves to the $x$-$y$ plane. The only remaining parameter determining the system behavior is now $v_0=\beta F\sqrt{D_t/D_r}$, which measures the relative strength of the self-propelling drive. From the data listed in Ref.~\cite{volpe2011microswimmers} we infer that $v_0=30$ constitutes a reasonable order of magnitude. 

In our case, the potential $V(\mathbf{r})$ only contains the effect of the cavity boundaries. 
Within the bulk of the cavity, the potential vanishes and the particle is completely free. 
Furthermore, $V$ does not act on the 
orientation of the self-propulsion direction $\mathbf{\hat{u}}$. This direction is solely determined by the independent rotational diffusion process Eq.~(\ref{rotdiff_phi}). It has been demonstrated in various works that, as a consequence, the global and long-time translational behavior of such particles is itself diffusive, with an effective diffusion constant increased by self-propulsion \cite{howse2007self,jiang2010active,volpe2011microswimmers,hagen2011brownian}. Despite these facts, we identify in the following a procedure that breaks the irreversibility implied by this long-time diffusive behavior, at least partially. 
\revision{Remarkably, our procedure 
is mediated by and only becomes possible due to rotational diffusion as described by Eq.~(\ref{rotdiff_phi}).} 

\section{Channel}\label{channel}

We start by considering an infinitely extended channel. 
This implies two parallel and infinitely extended confining walls, in our case along the $\mathbf{\hat{y}}$ direction. For simplicity, we choose the 
confining potential as 
\begin{equation}\label{Vchannel}
  V(\mathbf{r})=V(x)=\left\{\begin{array}{cl}
    k\left(|x|-\frac{L}{2}\right)^6 & \mbox{ if }|x|>\frac{L}{2},\\[.1cm]
    0                               & \mbox{ if }|x|<\frac{L}{2},
  \end{array}\right.
\end{equation}
where the width of the channel between the wall regions is given by $L>0$. The precise magnitude of the exponent and of the potential strength $k>0$ is not qualitatively important for the presented results. We set $k=50$. 

As an initial condition, we consider self-propelled particles located in the center of the channel with equally distributed orientations $\mathbf{\hat{u}}$. The corresponding probability density 
is independent of the $y$ coordinate, ${\nolinebreak\psi(x,\phi,t=0)}=\delta(x)/2\pi$. Consequently, the problem becomes translationally invariant in the $\mathbf{\hat{y}}$ direction. 
The spatial probability profile $\psi_x(x,t)$ across the channel is obtained by integrating out the $\phi$-dependency of $\psi(x,\phi,t)$. 
We choose a channel width of $L=20$, 
which in the framework of Ref.~\cite{volpe2011microswimmers} corresponds to about $11$ particle diameters. 

\begin{figure}
  \begin{center}
    \includegraphics[width=8.cm]{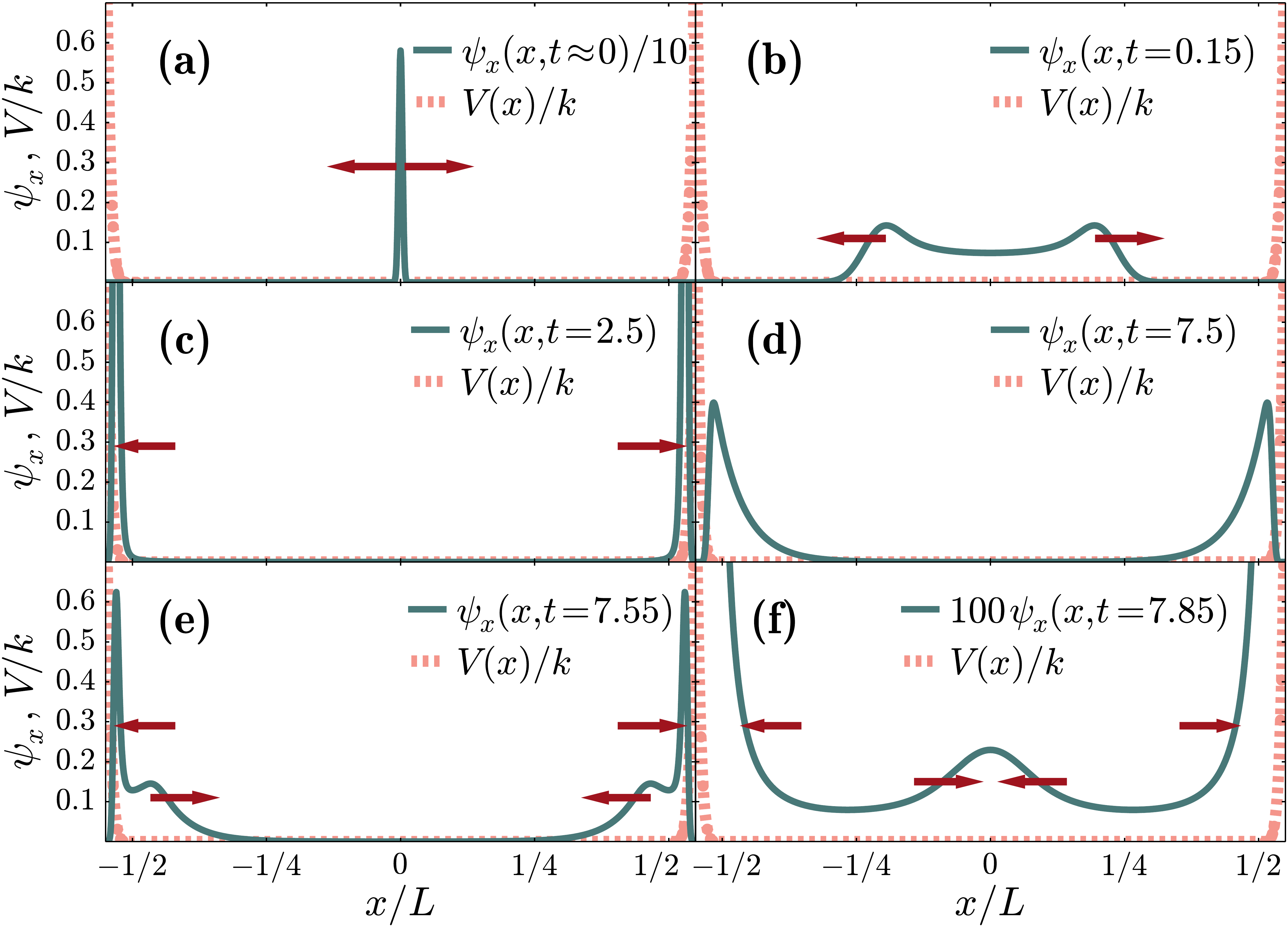}
    \caption{
    Time evolution of the spatial probability density distribution $\psi_x(x,t)$ across the channel of width $L$, bounded by a confining potential $V(x)$ of strength $k$. From the initial density peak (a) fronts migrate outward (b) and get blocked at the confining boundaries (c), where stationary density peaks build up. Switching off the self-propelling drive, these peaks broaden due to spatial diffusion (d); reorientation is possible due to rotational diffusion. Turning on the drive again (e), the peaks at the boundaries rebuild, but also two further peaks start to head back towards the channel center. When they overlay in the channel center (f), the initial spatial density distribution is partially restored. From here, the cycle can be repeated arbitrarily often without further losses. Dominating self-propulsion directions are marked by arrows. Parameter values are $L=20$, $v_0=30$, and $k=50$ in rescaled units. 
    } 
  \label{channel_figx}
  \end{center}
\end{figure}

Our protocol is the following. We numerically integrate the resulting dynamic equation for $\psi(x,\phi,t)$ forward in time. 
Fig.~\ref{channel_figx}~(a) shows the sharply-peaked initial probability density profile $\psi_x(x,t\approx0)$ at a very early stage, together with the confining potential $V(x)$. 
\begin{figure}
  \begin{center}
    \includegraphics[width=7.5cm]{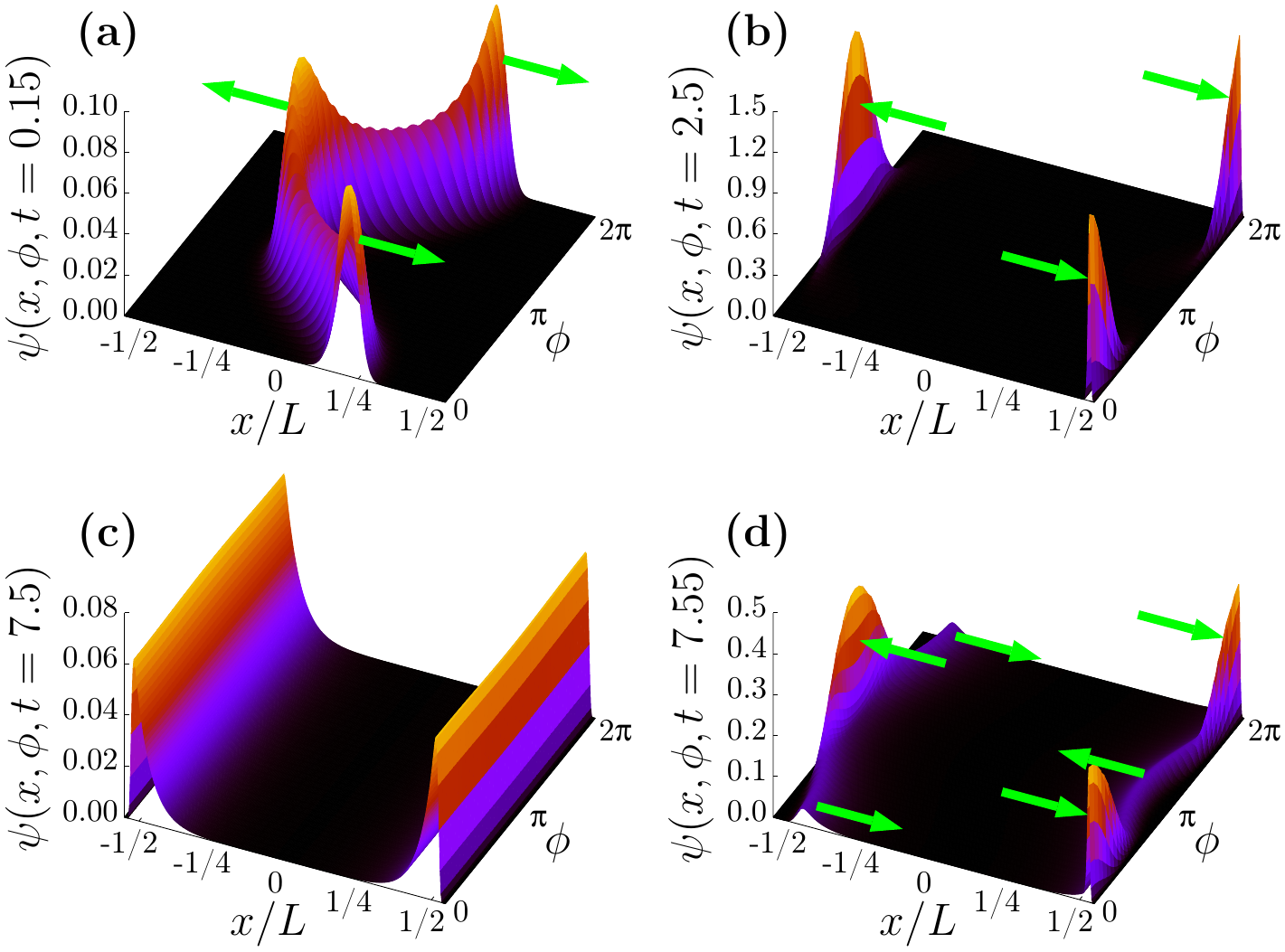}
    \caption{
Detailed probability density distribution $\psi(x,\phi,t)$ corresponding to Fig.~\ref{channel_figx}~(b)--(e) in the $x$-$\phi$ plane. At early times (a), the initial front at $x=0$ has split, leading to a peak at $\phi=0,2\pi$ that travels to the right, while the one at $\phi=\pi$ migrates to the left. The peaks get blocked and build up at the confining boundaries (b), centered around the self-propulsion directions that point outwards. When the self-propelling drive is switched off (c), these peaks broaden, leading to fronts of equally distributed orientations $\phi$. Finally, when self-propulsion is turned on again (d), those parts of the fronts corresponding to outward orientations reform the blocked peaks of (b); those parts of the fronts containing inward orientations start to propagate towards the channel center. Arrows mark the inward and outward self-propulsion directions. 
    } 
  \label{channel_figxphi}
  \end{center}
\end{figure}
Due to self-convection and since all orientations of the self-propulsion direction are equally probable at $t=0$, this peak splits and two fronts propagate outward towards the channel boundaries, see Fig.~\ref{channel_figx}~(b). At the boundaries, the probability density piles up and after a while reaches a stationary profile as displayed in Fig.~\ref{channel_figx}~(c). Such a blocking effect due to a confining potential has been observed in various forms before \cite{wensink2008aggregation,volpe2011microswimmers,elgeti2013wall,hennes2014self}: the self-propelling drive pushes the particles towards the walls, and the only way of escape is a reorientation of the self-propulsion direction. In our case, this reorientation is 
solely due to rotational diffusion. 

We now switch off self-propulsion, setting $v_0=0$. 
As mentioned above, for light-activated mechanisms this can simply be achieved by turning off the external illumination \cite{jiang2010active,volpe2011microswimmers,buttinoni2012active, palacci2013living,buttinoni2013dynamical,babel2014swimming}. 
This has two effects as displayed in Fig.~\ref{channel_figx}~(d). First, the potential pushes the particles back towards the edges $x=\pm L/2$. And second, translational diffusion 
leads to a broadening of the density peaks. Nevertheless, rotational diffusion is still active and generates within this localized particle cloud equally strong subpopulations of inward and outward oriented particles. 

When then self-propulsion is switched on again, $v_0\neq0$, the peaks at the boundaries split. Outward oriented particles return towards the channel boundaries. However, inward oriented particles form peaks that start to propel back towards the channel center, see Fig.~\ref{channel_figx}~(e). When they meet in the channel center, they overlay and again form a (smaller) central peak, as shown in Fig.~\ref{channel_figx}~(f). Consequently, the initial distribution \revision{has been restored partially, with significantly lower magnitude of the central peak (we show below that the restored peak height can be notably increased by steric interactions). Afterwards, following the same protocol, this central peak can be restored arbitrarily often without further losses. The restoration is possible} despite the irreversible diffusive processes involved. In fact, we have explicitly exploited a diffusive process to achieve our goal, namely rotational diffusion during the period of switched-off self-propulsion. 

Fig.~\ref{channel_figxphi} shows the whole process in the $x$-$\phi$ plane. There, we explicitly observe the initial splitting into outward propagating peaks centered around $\phi=0,2\pi$ and $\phi=\pi$, see Fig.~\ref{channel_figxphi}~(a). The outward oriented particles get trapped at the boundaries, leading to the peaks localized both in $x$ and $\phi$ direction in Fig.~\ref{channel_figxphi}~(b). After switching off the drive, rotational diffusion leads to equally $\phi$-distributed localized states at the boundaries. In other words, the peaks broaden to walls extended in the $\phi$ direction in Fig.~\ref{channel_figxphi}~(c). Finally, when the drive is turned on again, the inward oriented parts of these walls start to propagate back to the channel center, while the outward oriented ones return towards the wall regions, see Fig.~\ref{channel_figxphi}~(d). 

During this cycle, it is necessary to switch off the drive to generate with sufficient probability particle orientations towards the channel center. Turning on the drive, such particles immediately head towards the center, forming the propagating fronts that rebuild the central peak. 
If the drive were not switched off, a particle at a wall would always leave the boundary as soon as it slightly turns away. At this moment, due to the typically gradual rotational diffusion, the main component of the drive points along the bounding surface and not towards the channel center; pronounced inward-propagating fronts could not form. 

The time scales for the individual processes are determined by the intrinsic system parameters. It takes the ballistic time $L/2v_0\approx0.33$ for the edge of the outward propagating fronts in Fig.~\ref{channel_figx}~(b) to reach the channel boundaries. One should wait an order of magnitude longer for the stationary peaks in Fig.~\ref{channel_figx}~(c) to build up. Then, after switching off the drive, rotational diffusion over a time span $\Delta t$ leads to a mean squared angular displacement of $\langle(\Delta\phi)^2\rangle=2\Delta t$ in rescaled units. To achieve $\sqrt{\langle(\Delta\phi)^2\rangle}\gtrsim\pi$ to smoothen out the orientational distribution, a period of $\Delta t\gtrsim\pi^2/2\approx5$ is necessary. 
Yet, one should not wait too long so that the peaks are not smeared out by translational diffusion, see Fig.~\ref{channel_figx}~(d). Roughly estimating the mean squared displacement for the one-sided diffusion by $\langle(\Delta x)^2\rangle\sim\Delta t$ in rescaled units and requiring $\sqrt{\langle(\Delta x)^2\rangle}\lesssim L/8$, we obtain an upper constraint for the waiting time $\Delta t\lesssim L^2/64\approx6$. After turning on the drive again, the ballistic time $L/2v_0\approx0.33$ determines when the peak in the center of the channel reforms in Fig.~\ref{channel_figx}~(f). 
To observe the formation of the central peak, the channel center must be within reach of the self-propelling front before it is diluted by diffusion, i.e.\ $L/2\lesssim v_0$.

\begin{figure}
  \begin{center}
    \includegraphics[width=8.cm]{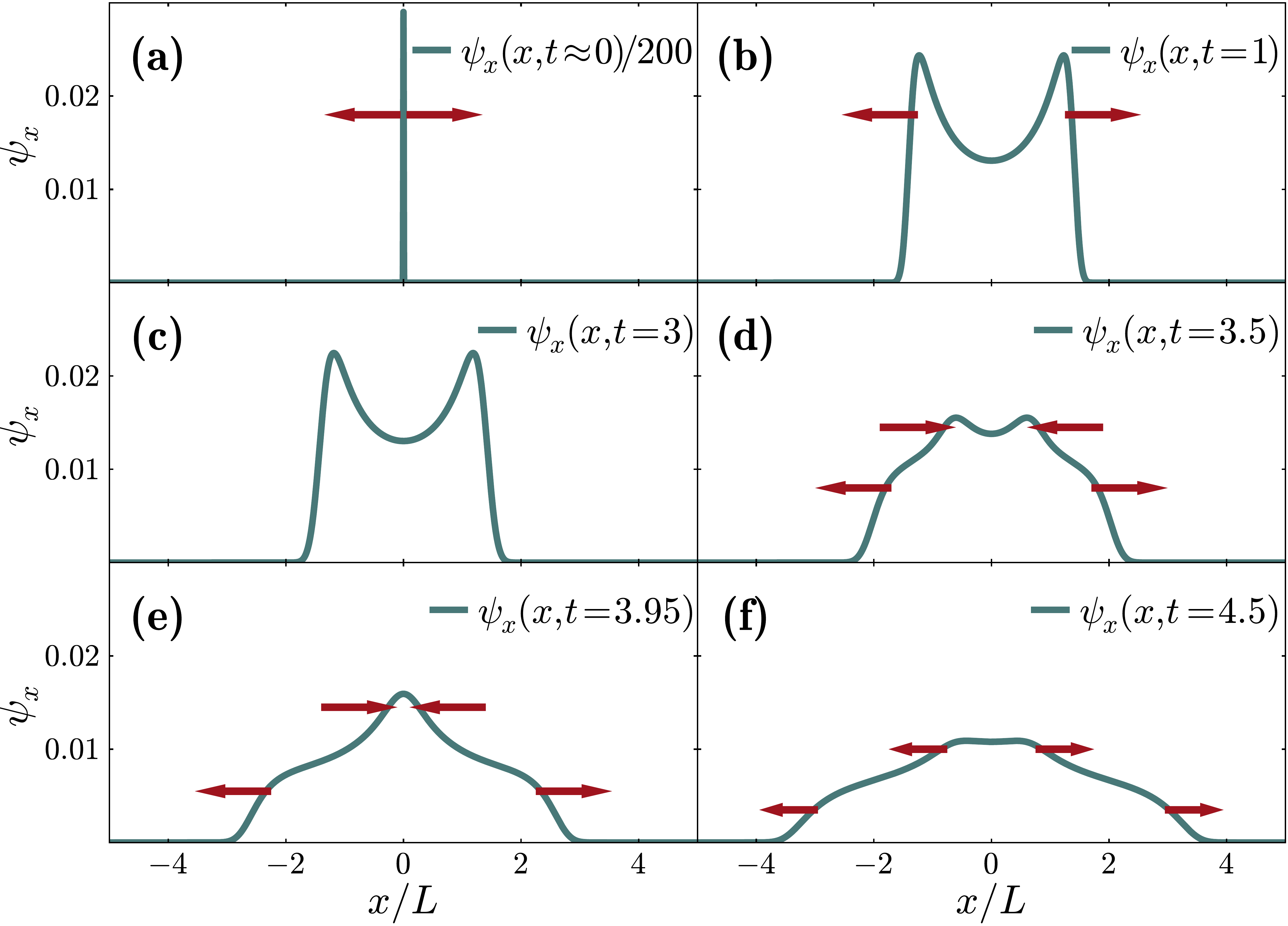}
    \caption{
    Same process as in Fig.~\ref{channel_figx}, but now without the confining channel walls, i.e.\ $V(x)\equiv0$. Spatial distances are rescaled by the same length $L$ for comparison. From the initial density peak (a) fronts migrate outward (b). Switching off the self-propelling drive, these free-standing peaks broaden due to spatial diffusion (c); reorientation is possible due to rotational diffusion. Turning on the drive again (d), the peaks split into two counterpropagating fronts. When the inward-traveling peaks overlay in the center (e), the initial spatial density distribution is partially restored. Over time, the density profile continuously flattens (f) due to the absence of confining boundaries. 
    } 
  \label{nochannel_figx}
  \end{center}
\end{figure}

We should mention that a similar process could also be achieved without boundaries. In Fig.~\ref{nochannel_figx} we switch off the drive after the initial separation, let the rotational diffusion work, and then switch on the drive again. 
This likewise leads to an overlay and partially restored peak at the initial starting point. However, over time, the peaks will more and more broaden due to diffusive processes, and density is lost towards the open boundaries. 
In contrast to that, a stationary state is involved in the form of the built-up peaks at the boundaries in Fig.~\ref{channel_figx}~(c). From here, the central peak in Fig.~\ref{channel_figx}~(f) can in principle be generated in an infinite number of repeatable cycles without further loss of restored peak intensity. 
The process becomes independent of initial conditions. 
Intervals of self-propulsion $v_0\neq0$ are necessary for this process to be observed.

%
%
%
\section{Circular geometry}\label{circular}

In an experiment using a channel structure, the channel ends may influence the statistics. Therefore, we additionally turn to a finite confinement. 
We consider a two-dimensional circular cavity of the same diameter $L$ as the channel width above. 
\begin{figure}
  \begin{center}
    \includegraphics[width=8.cm]{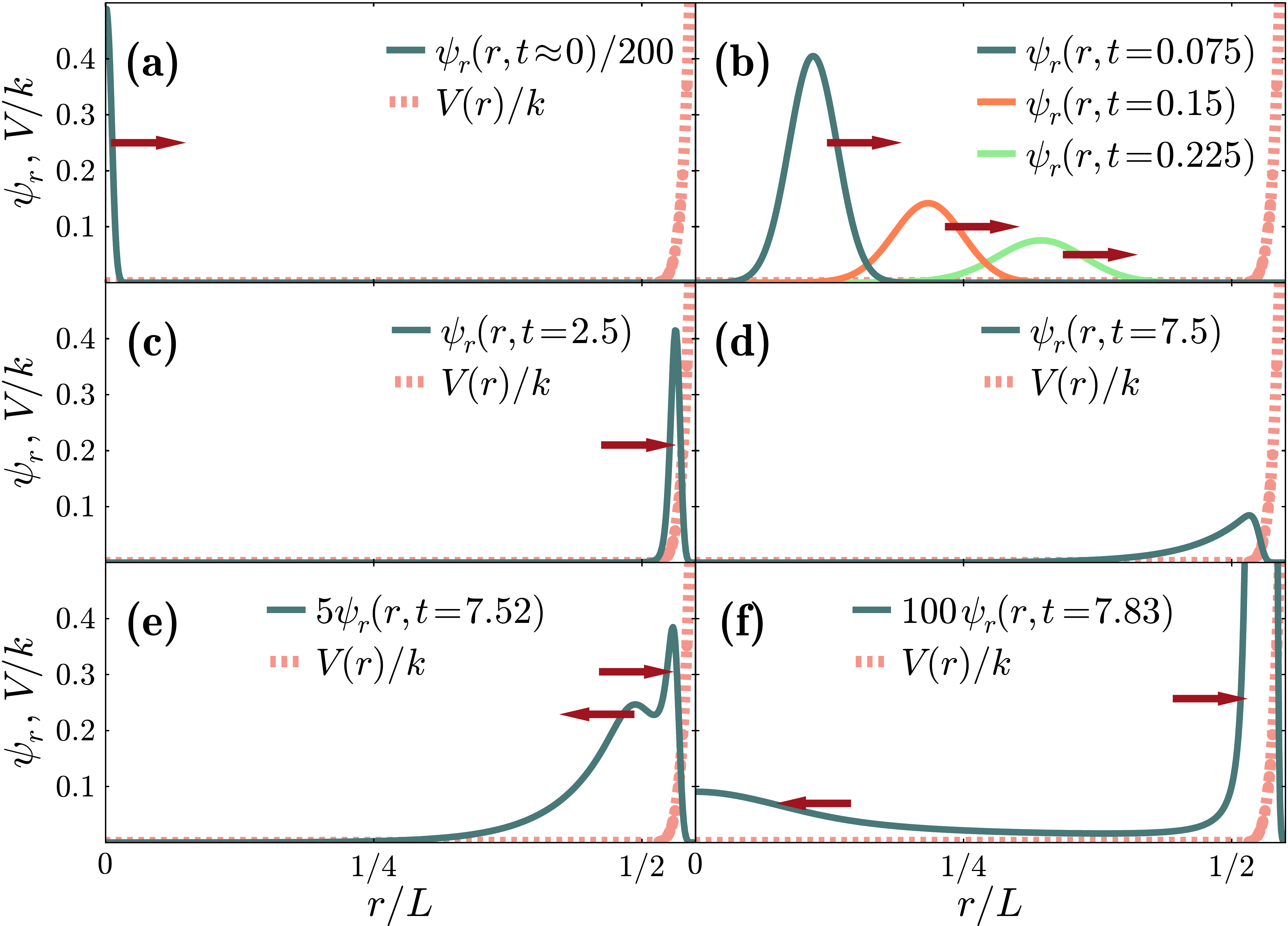}
    \caption{
    Time evolution of the spatial probability density distribution $\psi_r(r,t)$ in the circular cavity of diameter $L$, bounded by a confining potential $V(r)$ of strength $k$. From the initial central density peak at $r=0$ (a) a front migrates outward (b). It gets blocked at the cavity boundary (c), where a stationary density peak builds up. When the self-propelling drive is switched off, this peak broadens due to spatial diffusion (d); simultaneously, the orientations become equally distributed within the peak due to rotational diffusion. Turning on the drive again (e), the peak at the boundary rerises; an additional second peak emerges and heads back towards the cavity center. When it arrives there (f), the initial spatial density distribution is partially restored. From here, the cycle can be repeated arbitrarily often without further losses. Dominating self-propulsion directions are marked by arrows. Parameter values are $L=20$, $v_0=30$, and $k=50$ in rescaled units. 
    } 
  \label{circular_fig}
  \end{center}
\end{figure}
The confining potential 
$V(r)$, with $r=\|\mathbf{r}\|$, has the same form as $V(x)$ in Eq.~(\ref{Vchannel}). 
We assume an 
initial density distribution peaked in the center of the cavity at $r=0$, again with equally distributed orientations of the self-propulsion direction $\mathbf{\hat{u}}$. 
Thus the whole process becomes radially symmetric. We can reduce the number of variables measuring the self-propulsion direction by an angle $\vartheta$ relatively to the radial direction. 
For the resulting probability density $\psi(r,\vartheta,t)$ we obtain 
\begin{eqnarray}
\partial_t\psi & \hspace{-.0cm}=\hspace{-.1cm} & 
{}-v_0\Big[\cos\vartheta\:\partial_r\psi-\frac{1}{r}\sin\vartheta\:\partial_{\vartheta}\psi\Big] \nonumber\\
&&{}
+ \frac{1}{r}\partial_r\!\left[r\partial_r\psi\right] 
+ \left[1+\frac{1}{r^2}\right]\!\partial_{\vartheta}^2\psi
+\frac{1}{r}\partial_r \!\left[r(\partial_r\! V)\psi\right]. \hspace{.5cm}
\label{fp_rtheta}
\end{eqnarray}

We now follow the same protocol as before. 
The process is displayed in Fig.~\ref{circular_fig} for the radial probability density $\psi_r(r,t)$ obtained by integrating out the angle $\vartheta$ in $\psi(r,\vartheta,t)$. 
Starting from a density peak around $r=0$ in Fig.~\ref{circular_fig}~(a), particles migrating outward from the center of the cavity as in Fig.~\ref{circular_fig}~(b) get trapped at the boundary. There, a localized peak of elevated probability density forms, see Fig.~\ref{circular_fig}~(c). Switching off the self-propelling drive, rotational and spatial diffusion remain active and lead to equally distributed orientations within the broadening density peak, see Fig.~\ref{circular_fig}~(d). When the self-propelling drive is turned on again in Fig.~\ref{circular_fig}~(e) inward-pointing particles start to head towards the cavity center. After a while, a density peak is observed in the center of the confinement in Fig.~\ref{circular_fig}~(f). As before, the initial distribution has been partially restored. 
Furthermore, the situation in Fig.~\ref{circular_fig}~(f) can afterwards be completely restored arbitrarily often and does not depend on the initial condition any more. This is because an intermediate stationary state was involved in 
Fig.~\ref{circular_fig}~(c). 

\section{Steric interactions}\label{steric}

In Refs.~\cite{jiang2010active,volpe2011microswimmers,buttinoni2012active}, statistical properties of isolated light-switchable self-propelled particles were experimentally determined for dilute systems. 
We now briefly discuss what happens to the above results when particle interactions start to become important. 

The nature of steric \cite{bialke2013microscopic,speck2014effective} as well as of possible hydrodynamic interactions \cite{alarcon2013spontaneous,zottl2014hydrodynamics} depends on details of the particular system under consideration. To remain at this stage as general as possible, we only consider the most central feature of hard steric pair interactions: a diverging interaction energy when two particles are found at the same position. The simplest way to include this general property is by a steric pair interaction of the shape $v(\mathbf{r},\mathbf{r}')=\epsilon\delta(\mathbf{r}-\mathbf{r}')$, where $\mathbf{r}$ and $\mathbf{r}'$ mark the positions of the two particles and $\epsilon$ is the rescaled interaction strength. In the derivation of the statistical equations, we apply the mean-field approximation for the two-particle density $\psi^{(2)}$ by inserting $\psi^{(2)}(\mathbf{r},\phi,\mathbf{r}',\phi',t)=\psi(\mathbf{r},\phi,t)\psi(\mathbf{r}',\phi',t)$ \cite{menzel2011collective}. 
This leads to an additional contribution $\epsilon\nabla\!\cdot\!\left[\psi(\mathbf{r},\phi)\nabla\!\int\!\psi(\mathbf{r},\phi')d\phi'\right]$ in Eq.~(\ref{fp}). 
\begin{figure}
  \begin{center}
    \includegraphics[width=8.cm]{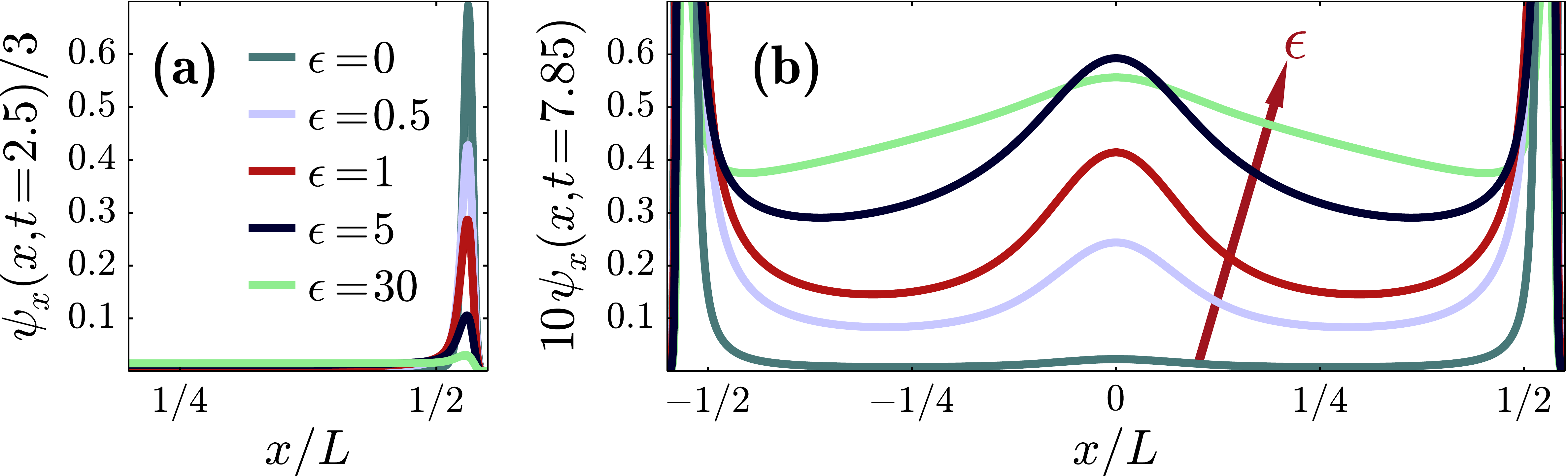}
    \caption{
    Same system as in Fig.~\ref{channel_figx} for the channel geometry, but now with an additional steric pair interaction $v(\mathbf{r},\mathbf{r}')=\epsilon\delta(\mathbf{r}-\mathbf{r}')$. For increasing interaction strength $\epsilon$, starting from $\epsilon\hspace{-1pt}=\hspace{-.5pt}0$, (a) the height of the stationary boundary peaks for $v_0\hspace{-1pt}\neq\hspace{-1pt}0$ decreases, whereas (b) the height of the restored central peak increases. The peaks are smeared out at elevated values of $\epsilon$. 
    } 
  \label{steric_channel}
  \end{center}
\end{figure}
\begin{figure}
  \begin{center}
    \includegraphics[width=8.cm]{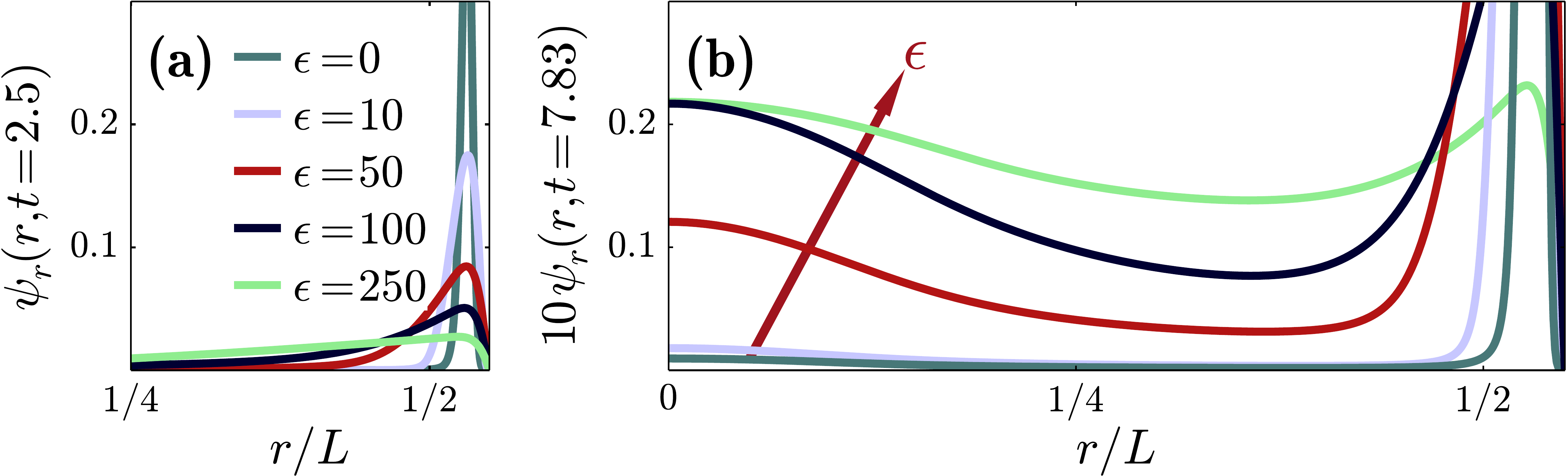}
    \caption{
    Same system as in Fig.~\ref{circular_fig} for the circular geometry, but now with a steric pair interaction $v(\mathbf{r},\mathbf{r}')=\epsilon\delta(\mathbf{r}-\mathbf{r}')$ included. Starting from $\epsilon=0$, (a) the height of the stationary boundary peaks for $v_0\neq0$ decreases, while (b) the height of the restored central peak increases for increasing interaction strength $\epsilon$. 
    At pronounced magnitudes of $\epsilon$ the peaks are smeared out. 
    } 
  \label{steric_circular}
  \end{center}
\end{figure}

There is now a certain pressure on the particles to leave highly populated areas. It is reflected by a lower magnitude of the stationary boundary peaks for $v_0\neq0$, see Figs.~\ref{steric_channel}~(a) and \ref{steric_circular}~(a) for the 
two considered geometries, respectively. 
Remarkably, close to the dilute limit, where the above approximations are restricted to, the steric interactions as a result enhance the build-up of the restored central peak, 
see Figs.~\ref{steric_channel}~(b) and \ref{steric_circular}~(b). 
At higher interaction strengths, 
the density distributions are smeared out.

\vspace{-.2cm}
\section{Conclusions}\label{conclusions}
\vspace{-.05cm}



\revision{
In this work, we have demonstrated how in a confined self-propelled particle system a central density peak can be repeatedly generated. Our procedure implies switching on and off the self-propulsion mechanism in combination with appropriate confining boundary interactions. If we start from an accordingly generated (or otherwise prepared) initial central density peak, this initial peak can (partially) be restored. After the first cycle of restoration, the central density peak can be repeatedly restored arbitrarily often without further losses in peak magnitude. In that sense, we break the irreversibility of this long-time diffusive system.}

Naturally, there are other examples where diffusive processes can be reversed. For instance, the diffusion of dipolar electric or magnetic particles can be counteracted by a guiding drift force resulting from an electric or magnetic field gradient. Yet, in such situations, the external forces directly pull the particles, imposing and pre-setting the drift directions. 
The present case is different. Here, each particle itself individually selects its migration direction in a stochastic process of rotational diffusion. 

Many experiments on self-propelling particles are performed in two-dimensional set-ups \cite{rappel1999self,nagai2005mode,szabo2006phase,narayan2007long, kudrolli2008swarming,goohpattader2009experimental, jiang2010active,thutupalli2011swarming,deseigne2010collective,volpe2011microswimmers, buttinoni2012active,peruani2012collective,theurkauff2012dynamic, zheng2013non,wada2013bidirectional,palacci2013living,buttinoni2013dynamical}. 
Accordingly, we here concentrated on two spatial dimensions. In principle, the analogous effect could also be analyzed in three dimensions. 
However, this renders the analysis considerably more tedious and requires a well-resolved three-dimensional detection technique on the experimental side. 

It should be straightforward to verify our results in corresponding experiments. Switching on and off the self-propulsion can easily be achieved for light-activated mechanisms \cite{jiang2010active,volpe2011microswimmers,buttinoni2012active, palacci2013living,buttinoni2013dynamical,babel2014swimming}. 
As briefly demonstrated, steric interactions among the particles can promote the effect. 
Details of interactions between particles \cite{buttinoni2013dynamical,elgeti2014physics,menzel2015tuned}, steric alignment interactions between non-spherical particles and cavity walls \cite{wensink2008aggregation,elgeti2009self,li2009accumulation, kaiser2013capturing,elgeti2014physics,menzel2015tuned}, or hydrodynamic interactions among the particles and with cavity walls \cite{elgeti2009self,lauga2009hydrodynamics,spagnolie2012hydrodynamics, hennes2014self,zottl2014hydrodynamics,elgeti2014physics,menzel2015tuned} may influence the statistics in different ways. The above theoretical description should then be adjusted to the specific situation under \pagebreak investigation. 

{\color{white}h}\\[-1.1cm]

\acknowledgments{
\vspace{-.08cm}
The author thanks the Deutsche Forschungsgemeinschaft (DFG) for support of this work through the priority program SPP 1726 ``Microswimmers'', grant no.~ME~3571/2-1. 
}


%

\end{document}